# Cognitive Biases in Software Engineering: A Systematic Mapping Study

Rahul Mohanani, Iflaah Salman, Burak Turhan, *Member, IEEE,* Pilar Rodríguez and Paul Ralph

**Abstract**—One source of software project challenges and failures is the systematic errors introduced by human cognitive biases. Although extensively explored in cognitive psychology, investigations concerning cognitive biases have only recently gained popularity in software engineering research. This paper therefore systematically maps, aggregates and synthesizes the literature on cognitive biases in software engineering to generate a comprehensive body of knowledge, understand state of the art research and provide guidelines for future research and practise. Focusing on bias antecedents, effects and mitigation techniques, we identified 65 articles (published between 1990 and 2016), which investigate 37 cognitive biases. Despite strong and increasing interest, the results reveal a scarcity of research on mitigation techniques and poor theoretical foundations in understanding and interpreting cognitive biases. Although bias-related research has generated many new insights in the software engineering community, specific bias mitigation techniques are still needed for software professionals to overcome the deleterious effects of cognitive biases on their work.

✦

## 1 INTRODUCTION

COGNITIVE biases are systematic deviations from optimal reasoning [1], [2]. In other words, they are recurring errors in thinking, or patterns of bad judgment observable in different people and contexts. A well-known example is *confirmation bias*—the tendency to pay more attention to information that agrees with our preconceptions.

Cognitive biases help to explain many common software engineering (SE) problems in diverse activities including design [3], [4], [5], testing [6], [7], requirements engineering [8], [9] and project management [10]. Research on cognitive biases is useful not only to identify common errors, their causes and how to avoid them in SE processes, but also for developing better practices [7], [8], methods [11], [12] and artifacts [3], [13].

While there is no definitive index, over 200 cognitive biases have been identified in psychology, sociology and management research. Some studies in the information systems (IS) community provide a general overview (e.g. [14], [15], [16]). In their systematic literature review, Fleischmann et al. [16] identified areas in IS where cognitive biases have already received substantial attention (e.g. IS usage, IS management) and areas where further investigation is warranted (e.g. software development, application systems). Such reviews help to synthesize and condense empirical findings into a more communicable, cumulative body of knowledge. No analogous review of SE research exists. The purpose of this study is therefore as follows:

> *Purpose: to review, summarize and synthesize the current state of software engineering research involving cognitive biases.*

Here, we use *current state of research* to encompass not only publication trends, publication outlets, research methods, SE knowledge areas and prolific authors but also bias antecedents, effects and mitigation approaches.

This study is timely for two reasons: 1) studies involving different cognitive biases tend to be disconnected from one another; 2) enough studies have now been published to facilitate substantial synthesis. This means that summarizing and aggregating research on cognitive biases in SE can now identify interesting patterns and produce new knowledge by connecting previously unconnected research.

This paper is organized as follows. The next section provides a brief primer on cognitive biases, including their causes, effects and consequences. Section 3 describes this study's research questions and method; that is, systematic mapping. We then present the findings in Section 4. Section 5 discusses the implications and limitations of our findings and offers avenues for future research. Section 6 concludes the paper with a summary of its contributions. Appendix A defines all of the biases discussed in this paper while Appendix B lists the primary studies.

## 2 COGNITIVE BIASES: A BRIEF PRIMER

Human behavior and cognition constitutes a critical research area in SE [17], [18], [19], [20]. Many studies on human decision-making in software engineering adopt theories and concepts from psychology to address issues in, for example, decision support systems [14], [21], software project management [11] and software engineering and development paradigms [22], [15], [23]. Several systematic

- *R. Mohanani, I. Salman and P. Rodríguez are with the M3S Group, University of Oulu, Finland.*
  E-mail: *rahul.mohanani@oulu.fi, iflaah.salman@oulu.fi and pilar.rodriguez@oulu.fi*
- *B. Turhan is with the Faculty of Information Technology, Monash University, Australia.*
  E-mail: *burak.turhan@monash.edu*
- *P. Ralph is with Department of Computer Science, University of Auckland, New Zealand.*
  E-mail: *paul@paulralph.name*

literature reviews have explored psychological and socio-logical aspects of SE including motivation [24], personality [25], organizational culture [26] and behavioral software engineering [27]. These studies collectively indicate the importance of human cognition and behavior for SE success.

While Section 4.3 provides a more comprehensive list, some examples of how cognitive biases affect SE include:

- *Confirmation bias* is implicated in the common antipattern where unit tests attempt to confirm that the code works rather than to reveal failures [28].
- *Optimism bias*, the tendency to produce unrealistically optimistic estimates [29], contributes to all kinds of projects (not only software projects) exceeding their schedules and budgets [30], [31].
- The *framing effect*—the tendency to give different responses to problems that have surface dissimilarities but are formally identical [2]—can reduce design creativity [3], [32].

Fleischmann [16] synthesizes a taxonomy of cognitive biases (based on previous studies [33]–[35]) comprising eight categories. Table 1 defines each of these categories. The categories have good face validity but it is not clear if they are exhaustive or mutually exclusive.

Identifying causes of cognitive biases is more challenging. Since Tversky and Kahneman [36] introduced the term in the early 1970s, a vast body of research has investigated cognitive biases in diverse disciplines including psychology [37], medicine [38] and management [16]. Much of this research has focused on heuristics—simple, efficient rules for making decisions and judgments. Well-known heuristics include anchoring and adjustment (i.e., estimating a quantity by adjusting from an initial starting value [39]) and availability (i.e., estimating the importance of something based on how easy it is to recall [40]).

Kahneman [33] went on to differentiate fast, "system one" thinking from slow, "system two" thinking. System one thinking is unconscious, effortless, intuitive, and prone to biases. System two thinking, contrastingly, is conscious, effortful, more insightful, more rational, more statistical and less prone to biases.

However, cognitive biases can be generated by phenomena other than system one thinking, including emotion [41], social influences [42] and noisy information processing [43]. The generative mechanisms for many cognitive biases remain poorly-understood.

SE researchers are also interested in *debiasing*; that is, preventing cognitive biases or mitigating their deleterious effects [23], [44]. As Fischoff [45] explains, debiasing interventions can be divided into five levels:

> To eliminate an unwanted behavior, one might use an escalation design, with steps reflecting increasing pessimism about the ease of perfecting human performance: (A) warning about the possibility of bias without specifying its nature; (B) describing the direction (and perhaps extent) of the bias that is typically observed; (C) providing a dose of feedback, personalizing the implications of the warning; (D) offering an extended program of training with feedback, coaching, and whatever else it takes to afford the respondent cognitive mastery of the task. (p. 426)

The problem with this approach is that options A, B and C are rarely effective [46], [47], and option D can be very expensive. Fischoff therefore offers a fifth option: debias the task instead of the person. Planning Poker, the effort-estimation technique used in many agile methods, illustrates option five—it redesigns effort estimation to prevent developers from anchoring on the first estimate [48].

However, a specific task like effort estimation can be corrupted by several interconnected biases [15]. Any particular redesign might not guard against all the biases in play. Planning Poker, for instance, attempts to mitigate anchoring bias but not optimism bias (cf. [49]).

## 3 RESEARCH OBJECTIVES AND METHOD

This section describes our research protocol (i.e. how we collected and analyzed data). We adopted a systematic mapping study (SMS) approach because we expected existing research to be fragmented and not follow common terminology or use established theoretical concepts. An SMS approach works well in such circumstances because it involves categorizing and aggregating knowledge dispersed across disconnected studies. This section details our approach, which is based on established guidelines (cf. [50]–[53]).

### 3.1 Research questions

Based on our research purpose—*to review, summarize and synthesize the current state of software engineering research involving cognitive biases*—the research questions for this systematic mapping study are as follows:

RQ1) What cognitive biases are investigated?
RQ2) What antecedents of cognitive biases are investigated?
RQ3) What effects of cognitive biases are investigated?
RQ4) What debiasing approaches are investigated?
RQ5) What research methods are used?
RQ6) When were the articles published?
RQ7) Where were the articles published?
RQ8) Who is conducting the research?
RQ9) Which SWEBOK knowledge areas are targeted by research on cognitive biases?

### 3.2 Protocol Overview

All the authors planned and agreed the study protocol visualized in Fig. 1. It consists of the following steps.

1) Search for *software AND "cognitive bias"* for all years up to and including 2016 in the selected online databases (see Section 3.3.1). (Retrieved 826 articles.)
2) Remove duplicates using automated then manual de-duplication (see Section 3.4.1). (Removed 308 duplicates.)
3) Apply the inclusion and exclusion criteria described in Section 3.4.2. (Included forty primary studies.)
4) Apply recursive backward snowballing exhaustively on reference lists to find additional studies (Section 3.5.1). (Added sixteen primary studies.)
5) Review the publication lists of the most active authors to find additional studies (Section 3.5.2). (Added nine primary studies.)

TABLE 1
Cognitive bias categories and descriptions (adapted from Fleischmann et al.'s [16] taxonomy)

| Bias category | Description | Examples |
|---|---|---|
| Interest biases | Biases that distort reasoning based on an "individual's preferences, ideas, or sympathy for other people or arguments" [16, p. 5]. | Confirmation bias, wishful thinking |
| Stability biases | Biases that lead individuals to persevere with established or familiar options despite the presence of superior information, arguments or conditions. | Anchoring bias, default bias |
| Action-oriented biases | Biases that lead individuals to premature decisions without considering relevant information or alternative courses of action. | Overconfidence bias, optimism bias |
| Pattern recognition biases | Biases that lead individuals to attend more to information that is familiar. | Availability bias, fixation |
| Perception biases | Biases that prejudice the processing of new information. | Framing effect, selective perception |
| Memory biases | Biases that affect recall of information or experience. | Hindsight bias, time-based bias |
| Decision biases | Biases that occur specifically during decision-making, and compromise decision quality. | Hyperbolic discounting, sunk cost bias |
| Social biases | Biases that prejudice judgment based on individuals' attitudes toward, and social relationships with, other individuals. | Bandwagon effect, cultural bias |

6) Check for sampling bias by searching for *software AND "cognitive bias"* on Google Scholar[1] as described in Section 3.5.3. (No primary studies added.)

7) Check for sampling bias by searching for *software AND [name of cognitive bias]*, using a list of cognitive biases (see Section 3.5.4 for more details). (Added one primary study for a total of 66.)

8) Assess the quality of the primary studies (see Section 3.6). (Excluded one study for a total of 65.)

9) Extract from the primary studies the data needed to answer the research questions (see Section 3.7).

## 3.3 Literature search

### 3.3.1 Database selection and search string formatting

We selected five online databases—IEEE Xplore, Scopus, Web of Science, the ACM Digital Library and Science Direct—following Dybå et al. [54]. These databases are frequently used to conduct secondary reviews on computer science and SE topics, and, in our experience, provide good coverage, filtering and export functionality. We used full-text search (rather than limiting to title and abstract). Where available, we used filters to avoid articles from outside SE.

To maximize coverage, we first tried the search string *software AND bias*. This produced an unmanageable number of articles (224037) including topics such as 'researcher bias' and 'publication bias', which are not *cognitive* biases. Replacing 'bias' with related terms including 'heuristics', 'fallacies', and 'illusions' similarly returned copious irrelevant results.

We therefore revised the search string to *software AND "cognitive bias"*. This produced a more manageable 826 records (Table 2). All searches were completed on 20 December 2016.

Before conducting these searches, we identified ten well-known, relevant studies based on our experience. We reasoned that, if any of these studies were missing from the sample, it would suggest a problem with the search strategy. All ten studies were found.

## 3.4 Primary study selection

### 3.4.1 De-duplication

We used RefWorks[2], to store and organize the retrieved articles, and to automatically remove 239 exact duplicates. We then exported the bibliographic entries to a spreadsheet and manually removed 69 additional duplicates.

### 3.4.2 Screening

Next, we applied the following inclusion criteria.

1) The article is written in English.
2) The context of the study is software engineering: "a systematic approach to the analysis, design, assessment, implementation, test, maintenance and re-engineering of software", that is, the application of engineering to software" [55].
3) The article explicitly involves at least one cognitive bias.
4) The article is published in a scholarly journal or in a conference, workshop or symposium proceedings.

We screened the studies initially based on their titles, consulting the abstract or entire text only when necessary to reach a confident judgment.

To develop a common understanding of the topic of study and inclusion criteria, two of the authors independently piloted the screening process on 20 randomly selected papers. This produced medium to high agreement (Cohen's Kappa = 0.77) [56]. Disagreements were resolved by consensus and discussion with the rest of the team, improving our common understanding of the inclusion criteria. This process raised concerns regarding non-peer-reviewed articles, academic dissertations, short papers, academic research plans, and studies where cognitive biases were mentioned only tangentially.

We therefore devised the following exclusion criteria.

1) The article is not peer-reviewed.

---

1. scholar.google.com      2. www.refworks.com

TABLE 2
Database Search Results

| Database | Search string | Filters applied | Results |
|---|---|---|---|
| Scopus | software AND "cognitive bias" | Included only the 'computer science' category. | 296 |
| Science Direct | software AND "cognitive bias" | Included only the 'journals' and 'computer science' categories. | 293 |
| IEEE Xplore | (software AND "cognitive bias") | Included only 'conference publications' and 'journals & magazines'. | 153 |
| ACM Digital Library | "software" AND "cognitive bias" | Excluded 'newsletters'. | 69 |
| Web Of Science | software AND "cognitive bias" | No filters were used. | 15 |

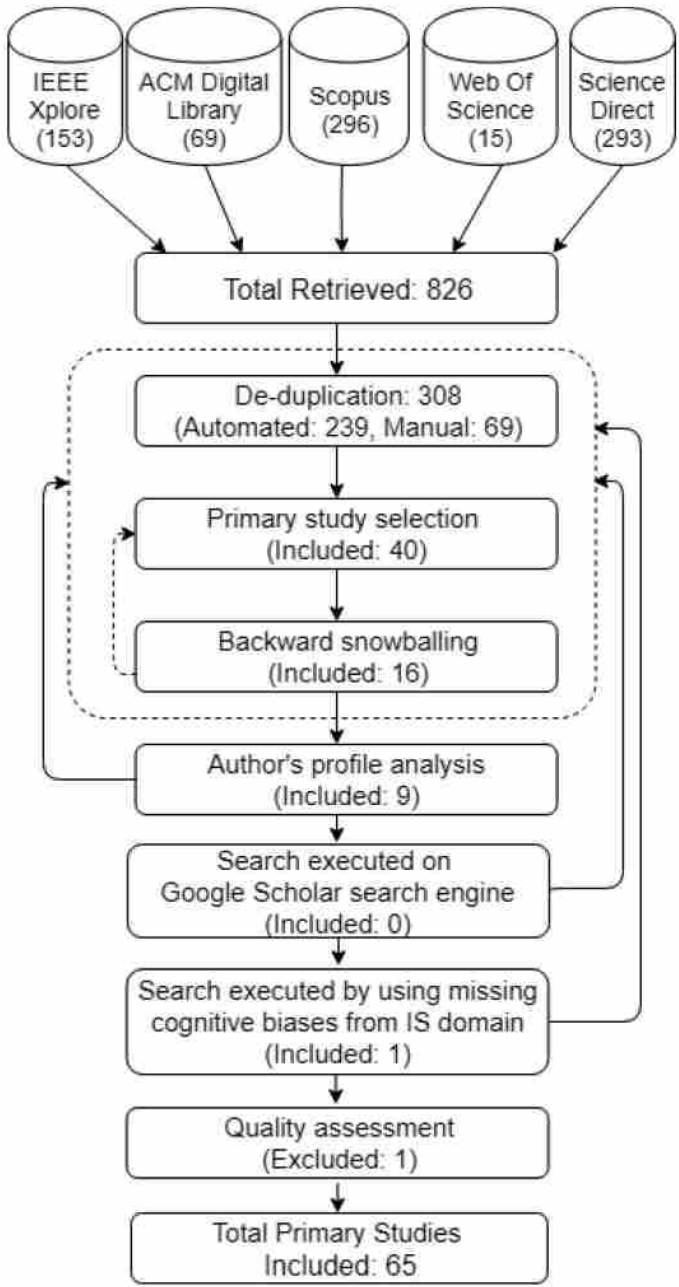

Fig. 1. Primary Study Selection Process

2) The article is a doctoral/master's dissertation or research plan, short paper or report thereof.

3) The article's references to cognitive biases were tangential to its purpose or contributions.

The third exclusion criteria was needed to exclude studies that mentioned a specific bias or cognitive biases in general, but did not *use* the concept of cognitive biases in the study. Some articles, for example, did not talk about cognitive biases at all, but the term "cognitive bias" appeared in the reference list. Other papers discussed cognitive biases in their related work, but made no further use of the concept.

The same two authors applied the inclusion and (new) exclusion criteria to another 20 randomly selected papers, yielding high agreement (Cohen's Kappa = 0.9). We deemed this sufficient to proceed, and completed the remainder of the screening. At the end of this stage, we had included 40 primary studies.

### 3.5 Ensuring sampling adequacy

This study has two basic threats to sampling adequacy:

1) There exist relevant articles in the population that are not in our sampling frame; for example, studies of cognitive biases published in non-SE outlets that are not indexed by the target search engines, or excluded by the filters we used (see Section 3.3.1).

2) There exist relevant articles in the sampling frame that are not in the sample; for example, an article about confirmation bias that never uses the term "cognitive bias".

We took several steps to mitigate these threats, including backward snowballing, author analysis, searching without filters and searching specific bias names.

#### 3.5.1 Backward snowballing

We iteratively and exhaustively checked the references of all included studies for additional relevant studies. That is, each time we included a new study, we also checked its references, leading to four iterations (Table 3) and 16 additional primary studies.

Upon closer inspection, 14 of the studies identified this way did not use the term "cognitive bias" anywhere in the text, preferring the name of the individual bias. The other two studies were published in information systems and psychology outlets.

#### 3.5.2 Author's analysis

Next, we reviewed the publication records of the most prolific authors. We reviewed the Google Scholar profile and personal website of each author of five or more primary studies. This produced nine additional primary studies (all of which were previously missed because they did not include the term "cognitive bias"). Backward snowballing on these new articles did not produce any additional primary studies.

TABLE 3
Recursive Backward Snowballing

| Snowballing round | Articles included |
| --- | --- |
| Round 1 | 10 |
| Round 2 | 4 |
| Round 3 | 1 |
| Round 4 | 1 |
| **Total** | **16** |

### 3.5.3 Unfiltered search

We searched Google Scholar for *software AND "cognitive bias"* with publication dates up to and including 2016. This returned over 200 000 records. We manually screened the first 150 results and found no new relevant studies.

### 3.5.4 Searching specific bias names

Ideally, we could use a comprehensive list of cognitive biases to search each target database for specific bias names (e.g. *software AND "framing effect"*). However, with over 200 known biases and five databases, this constitutes over 1000 individual searches. Moreover, we have no credible comprehensive list. We therefore compared the list of cognitive biases found in primary studies to the list in Fleischmann et al.'s review of cognitive biases in information systems [16], and found 18 biases investigated in IS but not in any of the primary studies— *after-purchase rationalization, ambiguity effect, attribution error, base-rate fallacy, congruence bias, consistency bias, cultural bias, emotional bias, exponential forecast bias, input bias, irrational escalation, loss aversion bias, mental accounting, priming effect, reactance, selective perception, stereotype bias* and *zero-risk perception.*

We searched each database for each of 18 new biases (e.g. *software AND ("ambiguity effect" OR "emotional bias" OR ...)* ) and found one new study (66 in total). Backward snowballing on the reference list retrieved no new studies.

### 3.6 Quality assessment

To assess the quality of the primary studies, we synthesized multiple recommendations [54], [57], [58] into a single set of yes-or-no quality assessment criteria (Table 4). Since many of these criteria only apply to empirical studies, we divided the included studies into empirical (47) and non-empirical (19). Here, an *empirical* study is simply one that analyses primary data; *non-empirical* studies include both opinion papers and wholly conceptual research. Moreover, quality here refers to an article's reporting practices, rather than judging the methodological rigor of the study itself.

To improve rigor and reliability, we ran a pilot where four authors independently assessed the quality of ten randomly selected primary studies, discussed disagreements, refined mutual understanding of the categories, and revised the quality assessment instrument. Then, the first two authors each assessed half of the remaining studies.

The mean quality score (percentage of quality criteria satisfied against overall quality criteria check-list) for empirical studies is 84%; for non-empirical studies, 85%. We excluded one non-empirical study that met fewer than 50% of the quality criteria.

This brought the total to 65 primary studies: 47 empirical and eighteen non-empirical (Table 5). See Appendix B for a complete list.

### 3.7 Data extraction

Next, we extracted the data elements shown in Table 6. We used the qualitative data analysis software NVIVO[3], because it is good for organizing, analyzing and visualizing unstructured qualitative data, and for classifying articles based on bibliographic information. We coded parts of each primary study with the name of the cognitive bias investigated, wherever an antecedent, effect or debiasing approach is discussed. We then organized the data based on the research questions (see Section 4). The coding schema and classifications were thoroughly reviewed by the whole author team.

## 4 RESULTS

This section addresses each research question (RQ1–RQ9). We refer to primary studies with a 'P' (e.g. [P42]) to distinguish them from citations to other references (e.g. [42]).

### 4.1 What cognitive biases are investigated?

The primary studies investigate 37 different cognitive biases, which are listed and defined in Appendix A. Fig. 2 shows the frequency of occurrence of each cognitive bias broken down by SE knowledge area. The most common are anchoring bias (26), confirmation bias (23), and overconfidence bias (16).

The exact number of cognitive biases investigated is difficult to pin down because some articles use different names for the same cognitive bias. We combined obvious synonyms (e.g. "optimism bias" and "over-optimism bias"). We did not combine closely related biases (e.g. optimism bias and the valence effect). Section 5.2 returns to the issue of inconsistent terminology.

To get a higher-level view of what kinds of cognitive biases are used more or less frequently, we categorized the biases using Fleishmann's taxonomy [16] (Table 7). The total number of studies in Table 7 add up to more than the number of included primary studies (i.e. 65), as some primary studies investigate cognitive biases in multiple categories.

Two authors independently categorized the biases and resolved disagreements through discussion. As the categories may not be mutually exclusive, we classified each bias into the most appropriate category. Some biases (e.g. information bias) do not fit into any category. These results should be considered in light of mediocre 55% agreement between judges.

3. www.qsrinternational.com

TABLE 4
Quality Assessment Checklist

| Quality Criteria | Empirical | Non-empirical |
|---|---|---|
| Was a motivation for the study provided? | X | X |
| Was the aim (e.g. objectives, research goal, focus) reported? | X | X |
| Was the study's context (i.e. knowledge areas) mentioned? | X | X |
| Does the paper position itself within existing literature? | X | X |
| Is relevance (to industry OR academia) discussed? | X | X |
| Were the findings or conclusion reported? | X | X |
| Was the research design or method described? | X | |
| Was the sample or sampling strategy described? | X | |
| Was the data collection method(s) reported? | X | |
| Was the data analysis method(s) reported? | X | |
| Were limitations or threats to validity described? | X | |
| Was the relationship between researchers and participants mentioned? | X | |

TABLE 5
Classification of Primary Study Based on Research Type

| Type | Description | Primary studies | Total |
|---|---|---|---|
| Empirical | Articles that analyze *primary* data such as quantitative measurements, questionnaire responses, interviews, archival data and field notes. | P1, P2, P4, P5, P6, P8, P9, P10, P11, P12, P13, P15, P16, P17, P18, P19, P24, P25, P28, P29, P31, P34, P38, P39, P40, P41, P42, P43, P44, P46, P47, P48, P49, P50, P51, P52, P53, P54, P55, P56, P57, P58, P59, P61, P62, P63, P65 | 47 (72%) |
| Non-empirical | Articles that do not analyze primary data, including not only opinion papers but also analytical, conceptual and purely theoretical research. | P3, P7, P14, P20, P21, P22, P23, P26, P27, P30, P32, P33, P35, P36, P37, P45, P60, P64 | 18 (28%) |

TABLE 6
Data Extraction Elements

| Element | RQ | Description |
|---|---|---|
| Cognitive bias | RQ1 | The bias(es) investigated or discussed in the study. |
| Antecedents | RQ2 | Causes of biases investigated or discussed in the study. |
| Effects | RQ3 | Consequences of biases investigated or discussed in the study. |
| Debiasing approaches | RQ4 | Treatments, techniques, or strategies reported to prevent biases or mitigate their effects. |
| Research method | RQ5 | The kind of study (e.g. experiment, case study). |
| Year | RQ6 | The year in which the study was published. |
| Source | RQ7 | The journal, conference or workshop where the study was published or presented. |
| Authors' names | RQ8 | The authors of the study. |
| SE knowledge area | RQ9 | The knowledge area (from SWEBOK version 3) to which the study contributes. |

Biases from all eight categories are investigated. The most investigated categories are interest biases (which distort reasoning based on the individual's interests) and stability biases (which bias reasoning in favour of current conditions). In contrast, only two studies investigate social biases (in which social relationships undermine judgment).

## 4.2 What antecedents of cognitive biases are investigated?

Although the primary studies investigate 37 cognitive biases, they only consider the antecedents of eleven (Table 8). This section discusses each in turn.

### 4.2.1 Anchoring and adjustment bias

Anchoring and adjustment is a common heuristic in which one makes estimates by adjusting an initial value called an anchor. Anchoring bias is the tendency to stick too closely to the anchor [36].

Suppose, for example, that a team is distributed across two separate offices. Both teams have a large whiteboard for sprint planning. In the middle of office A's whiteboard it says "What can we finish in the next two weeks?", but office B's whiteboard says "two days." Now suppose the team has a virtual meeting to estimate the time needed to build a specific feature, which will take about a week. Anchoring bias explains team members in office A giving higher estimates than team members in office B.

Several primary studies (e.g. [P16], [P17]) suggest potential antecedents of anchoring bias, including inexperience. Jain et al. suggest that experts are less susceptible to anchoring bias than novices [P16]. Meanwhile, uncertainty, lack of business knowledge and inflexible clients exacerbate anchoring during decision-making in project-based organizations [P63]. Other biases, including confirmation bias and availability bias, may also promote anchoring and adjustment [P17].

Anchoring bias can be confusing because Tversky and Kahneman [36] proposed it in the context of numerical estimates, but the term has since broadened to include fixating on any kind of initial information. Several primary

TABLE 7
Primary studies organized into Fleischmann et al.'s [16] taxonomy

| Bias category | Cognitive biases | # of studies | Primary study ID |
|---|---|---|---|
| Interest | Confirmation, IKEA effect, valence effect, validity effect, wishful thinking | 30 | P4, P5, P6, P7, P8, P10, P12, P13, P14, P16, P17, P18, P20, P21, P22, P25, P26, P27, P30, P31, P32, P36, P37, P39, P43, P44, P49, P55, P57, P64. |
| Stability | Anchoring and adjustment, belief perseverance, default, endowment effect, status-quo | 29 | P1, P2, P7, P10, P11, P16, P17, P18, P20, P21, P22, P25, P30, P33, P34, P37, P38, P42, P43, P47, P49, P51, P53, P56, P57, P60, P63, P64, P65. |
| Action-oriented | (Over)confidence, impact, invincibility, miserly information processing, misleading information, normalcy effect, (over)optimism | 22 | P3, P16, P20, P21, P22, P24, P31, P37, P41, P45, P46, P48, P49, P50, P51, P52, P54, P57, P58, P59, P61, P62. |
| Pattern recognition | Availability, mere exposure effect, fixation, Semmelweis reflex | 19 | P7, P10, P11, P13, P16, P17, P18, P19, P22, P23, P29, P32, P37, P39, P43, P60, P63, P64, P65. |
| Perception | Attentional, contrast effect, framing effect, halo effect, primacy and recency effect, selective perception, semantic fallacy | 9 | P19, P20, P28, P35, P36, P40, P60, P63, P64. |
| Memory | Hindsight, time-based bias | 6 | P30, P36, P49, P57, P60, P63. |
| Decision | Hyperbolic discounting, infrastructure, neglect of probability, sunk cost | 3 | P36, P37, P63. |
| Social | Bandwagon effect | 2 | P21, P22. |
| Other | Representativeness, information | 9 | P7, P10, P15, P29, P32, P36, P43, P53, P60. |

studies (e.g. [P2], [P20], [P33]) use this broader meaning.

For example, one primary study found that system designers anchor on preconceived ideas, reducing their exploration of problems and solutions [P33]. Similarly, developers tend to re-use existing queries rather than writing new ones, even when the existing queries do not work in the new situation [P2]. Two primary studies ([P17], [P18]) investigated developers insufficiently adjusting systems due to change requests. They found that missing, weak, and neglect of traceability knowledge leads developers to anchor on their own understanding of the system, which causes poor adjustments (i.e., deficient modifications). Another study [P20] found that decision-makers struggled to adjust to new or changed environments because they were anchored on their outdated understanding of a previous context.

### 4.2.2 Optimism bias

Optimism bias is the tendency to produce overly optimistic estimates, judgments and predictions [59]. Most of the primary studies concerning optimism bias study effort estimation among practitioners. Several factors appear to aggravate optimism bias.

Practitioners in technical roles (e.g. developers) appear more optimistic than those in non-technical roles (e.g. managers), at least in estimating web development tasks [P50]. The implication here is surprising: knowing more about specific tasks can lead to less accurate effort estimates.

Meanwhile, one study of students enrolled in a software development course found that estimates were more accurate for larger tasks and in hindsight. Students are optimistic about the accuracy of their estimates (i.e., their confidence intervals are far too tight) and are more optimistic for more difficult tasks and more pessimistic for easy tasks [P52].

Human perceptions and abilities also appear to affect optimism. For example, practitioners judge more concrete, near-future risks more accurately than more abstract, far future risks [P24].

Optimism bias is also related to project initiation. For example, *winner's curse* is the phenomenon in which the most competitive—not the most realistic—bid is selected [P59].

### 4.2.3 Availability bias

Availability bias is the tendency for easy-to-recall information to unduly influence preconceptions or judgments [60].

For example, availability bias manifests in two ways when searching software documentation [P13]: (1) professionals use inappropriate keywords because they are familiar and easy to remember; (2) professionals struggle to find answers in unfamiliar locations in documentation. Nevertheless, prior knowledge has a positive overall effect on efficiency and effectiveness in documentation searches [P13].

Availability bias can also manifest in project-based organizations, when decisions are based on information that is easy to recall or in the absence of lessons learned from previously-completed projects [P63].

### 4.2.4 Confirmation bias

Confirmation bias is the tendency to pay undue attention to sources that confirm our existing beliefs while ignoring sources that challenge our beliefs [14]. Confirmation bias manifests in software testing as 'positive test bias' [61]. That is, developers tend to test their programs with data that is consistent with 'the way that the program is supposed to work', and not with any inconsistent data [P31]. Specification completeness, domain knowledge and the presence or absence of errors (e.g. run-time errors, logical errors, user-interface errors) moderate this bias, but programming expertise seems to have no effect [P31].

Another primary study suggests the presence of prior knowledge in professionals from past projects as an antecedent to confirmation bias. Professionals while seeking information from software documentation, tend to focus the search only on information that confirms their existing knowledge. Prior knowledge therefore has a negative net effect on information retrieval from software documentation [P13].

TABLE 8
Antecedents of cognitive biases

| Bias | Antecedents | Knowledge area |
|---|---|---|
| Anchoring and Adjustment | Reusing previously written queries; difficult to identify referential points (anchors) [P2] | Construction |
| | Missing, weak and disuse of traceability knowledge [P17], [P18] | Design, Construction |
| | Recalling domain related information from past knowledge [P18] | Construction |
| | Not being able to adjust to the new environment [P20] | Requirements |
| | Development experience [P16] | Construction |
| | Uncertainty of future actions, lack of business / historical knowledge and inflexible clients [P63] | Management |
| | Confirmation and availability bias during design; [P17] | Design |
| Optimism | Technical roles (project manager, technology developers); effort estimation strategy [P50] | Management |
| | Task difficulty; task size [P52] | Management |
| | Estimates asking format [P47] | Management |
| | Psychologically distant project risks [P24] | Management |
| | Winners curse in project bidding [P59] | Management |
| Availability | Prior-knowledge; earlier experience in searching [P13] | Management |
| | Lack of historical records of lessons learned [P63] | Management |
| | Selection of test-cases due to impossibility to perform exhaustive testing [P32] | General |
| Confirmation | Prior-knowledge [P13] | Maintenance |
| | Lack of training in logical reasoning and mathematical proof reading; experience and being active in a role (developer / tester) [P4] | Testing |
| | Varied level of domain-expertise, specification completeness, presence of errors [P31] | Testing |
| Fixation/ framing | Framing desiderata as "requirements" [P19] | Requirements |
| Overconfidence | Inability to question the fundamental way of thinking [P45] | Management |
| Hindsight | Weak historical basis of lessons learned [P63] | Management |
| Mere exposure | Assigning someone to a different project role, leaving one's comfort zone, pessimism about change outcomes [P63] | Management |
| Halo effect | Subjective evaluation criteria being affected by the first impression [P63] | Management |
| Sunk cost bias | Extending decisions based on immediate benefits without considering costs of changes; stubbornness and comfort zone characteristics of project managers [P63] | Management |

Interestingly, professionals who were experienced but inactive as developer/testers showed less confirmation bias than those who were both experienced and active [P4]. Participants who have been trained in logical reasoning and hypothesis testing skills manifested less confirmation bias in software testing [P4].

### 4.2.5 Fixation and framing effect

Fixation is the tendency to focus disproportionately on one aspect of a situation, object or event—particularly self-imposed or imaginary barriers [15]. Only one primary study explicitly investigated fixation; it found that framing desiderata as "requirements" leads to more fixation on those desiderata than framing them as a less-structured list of "ideas" [P19].

### 4.2.6 Overconfidence bias

Overconfidence bias is the tendency to overestimate one's skills and abilities [62]. One primary study investigates overconfidence bias. It argues that focusing on brief tasks might lead project managers to neglect other useful information, leading to illusion of control and overconfidence in their abilities [P45]. The lack of ability to reflect on one's own experiences might also lead to overconfidence bias among software developers [P45].

### 4.2.7 Hindsight bias, mere exposure effect, halo effect and sunk cost bias

One qualitative interview study [P63] includes numerous participant speculations as to the causes of several biases. To be clear, interviews do not demonstrate causal links, but we report the speculated causes for completeness.

Hindsight bias refers to the tendency to regard the outcome of a process as having been predictable all along, or at least more predictable than they would have judged before knowing the outcome [63]. The absence of archived records from previous projects may increase hindsight bias [P63].

The mere exposure effect is the tendency to prefer the familiar [64]. For example, project participants tend to prefer their current role in the project [P63].

The halo effect is the tendency to use global evaluations to make judgments about specific traits [65]. In a job interview, for example, the interviewer might incorrectly infer that more charismatic candidates are more technically competent. Several potential antecedents for the halo effect are suggested, including "subjective evaluation criteria ... communication problems, the focus on people faults, and the absence of transparency in sharing personal problems" (p. 5).

Sunk cost bias is the tendency to irrationally invest more resources in a situation, in which a prior investment has been made, as compared with a similar situation in which a prior investment has not been made [66]. Sunk cost bias may by rooted in personality traits of project managers, including

| Cognitive Bias \ Knowledge Area | Computing foundations | Configuration | Construction | Design | Economics | Engineering foundations | General | Maintenance | Management | Mathematical foundations | Models and Methods | Process | Professionalism | Quality | Requirements | Testing | Total |
|---|---|---|---|---|---|---|---|---|---|---|---|---|---|---|---|---|---|
| Anchoring/adjustment | | | 5 | 8 | | | | | 9 | | 1 | | | 1 | 2 | | 26 |
| Attentional | | 1 | | | | | | | | | | | | | | | 1 |
| Availability | | | 5 | 4 | | | | | 3 | | | 1 | | 1 | 1 | | 15 |
| Bandwagon effect | | | | 1 | | | | 1 | | | | | | | | | 2 |
| Belief perseverance | | | | 1 | | | | | 1 | | | | | | | | 2 |
| (Over)-confidence | | | | 1 | | | | | 15 | | | | | | | | 16 |
| Confirmation | | | 6 | 4 | | | 1 | 1 | | | | | | 1 | 3 | 7 | 23 |
| Contrast effect | | | | 1 | | | | | | | | | | | | | 1 |
| Default | | | | 1 | | | | | | | | | | | | | 1 |
| Endowment effect | | | | | | | | | | | | | | | 1 | | 1 |
| Fixation | | | | 1 | | | | | | | | | | | 1 | | 2 |
| Framing effect | | | | 1 | | | | | | | | | | | 2 | | 3 |
| Halo effect | | | 1 | | | | 1 | | | | | | | | | | 2 |
| Hindsight | | | | | | | | | 3 | | | | | | | | 3 |
| Hyperbolic discounting | | | | 1 | | | | | | | | | | | | | 1 |
| IKEA effect | | | | | | | | | | | | | | | 1 | | 1 |
| Impact | | | | | | | | | 2 | | | | | | | | 2 |
| Information | | | 2 | 1 | | | 1 | | 1 | | | | | | | | 5 |
| Infrastructure | | | | 1 | | | | | | | | | | | | | 1 |
| Invincibility | | | | 1 | | | | | | | | | | | | | 1 |
| Mere exposure effect | | | 1 | | | | | | 2 | | | | | | | | 3 |
| Miserly info. processing | | | | | | | | | | | | 1 | | | | | 1 |
| Misleading information | | | | 1 | | | 1 | | 2 | | | | | | | | 3 |
| Neglect of probability | | | | 1 | | | | | | | | | | | | | 1 |
| Normalcy effect | | | | 1 | | | | | | | | | | | | | 1 |
| (Over)-optimism | | | 1 | 1 | | | | | 1 | | | | | | 3 | | 6 |
| Primacy / recency | | | | 1 | | | | | | | | | | | | | 1 |
| Representativeness | | | | | | | | | 1 | 1 | | | | 1 | 1 | | 4 |
| Selective perception | | | | | | | 1 | | | | | | | | | | 1 |
| Semantic fallacy | | | 1 | | | | | | | | | | | | | | 1 |
| Semmelweis reflect | | | 1 | | | | | | | | | | | | | | 1 |
| Status quo | | | | 2 | | | | | 1 | | | | | | | | 3 |
| Sunk cost | | | | | | | | | 1 | | | | | | | | 1 |
| Time-based | | | | 1 | | | | | | | | | | | | | 1 |
| Valence effect | | | | 1 | | | | | | | | | | | | | 1 |
| Validity effect | | | | 1 | | | | | | | | | | | | | 1 |
| Wishful thinking | | | | 1 | | | | | 2 | | | | | | | | 3 |
| **Total** | 0 | 0 | 22 | 39 | 0 | 0 | 5 | 2 | 44 | 0 | 2 | 2 | 0 | 4 | 15 | 7 | |

Fig. 2. Cognitive Biases Investigated in SE Knowledge Areas

stubbornness and myopia.

## 4.3 What effects of cognitive biases are investigated?

Although the primary studies investigate 37 cognitive biases, they only consider the effects of eleven (Table 9). This section discusses each in turn.

### 4.3.1 Anchoring and adjustment bias

Anchoring and adjustment bias affects multiple aspects of software development including effort estimation, implementing change requests and prototyping.

In the context of artifact and query reuse, developers who anchor on existing solutions tend to include unnecessary functionality. Errors in the anchors tend to propagate into the new artifacts [P2], [P42].

Project managers often tend to anchor their initial time, cost and effort estimates to lower values. This distorts the assessment of software productivity, which makes it difficult to adjust project plans dynamically. For example, by anchoring on a very conservative (lower) project completion time estimate, project managers tend to overestimate software productivity. Software developers then feel pressured to work harder to bring the project back on schedule. Eventually, project managers acquire additional resources, exceeding the project's budget and schedule [P51].

In the context of implementing change requests, developers may anchor on their initial understanding of the desiderata, the design artifact, or an initial solution concept. Developers then tend to adjust these anchors insufficiently to accommodate new or conflicting desiderata [P17].

Anchoring and adjustment bias is also observable among students. For example, one primary study reported a pattern in which graduate students simultaneously completed courses in software architecture (in general) and service-oriented architecture. When completing in-class assignments in software architecture, the students appeared to anchor on their concurrent study of service-oriented architecture [P64]. However, the study does not comment on the quality of the designs or whether using service-oriented principles was particularly inappropriate.

### 4.3.2 Confirmation bias

The tendency to pay more attention to information that confirms our beliefs affects many areas of SE. For example, due to confirmation bias, software professionals tend to be unrealistically positive about the completeness and correctness of search-queries for retrieving information from archived software documentation. Professionals tend to only search familiar areas of documentation, confirming their existing views [P13].

Confirmation bias also affects trade off studies (a practice in which a multidisciplinary team evaluates alternative solutions against multiple criteria). Participants tend to ignore results that do not confirm their preconceived ideas [P37]. Similarly, developers tend to resist exploring major design changes necessary for accommodating change requests that include new or conflicting requirements [P18].

Meanwhile, confirmation bias in software testing increases production defects [P4], [P6]. Confirmation bias explains the tendency of software testers to design and run more tests that confirm the expected execution of a program rather than ones that could reveal the failures [P31], [P32]. During debugging, confirmation bias sometimes leads programmers to misidentify the reasons for program failure [P32].

### 4.3.3 Hyperbolic discounting

Hyperbolic discounting is the tendency to prefer smaller rewards in the near future over larger rewards later on [67]. One primary study argues hyperbolic discounting leads software professionals to prefer quick-fixes and simple refactoring over better but more ambitious revisions [P36].

### 4.3.4 Availability bias

Software professionals tend to rely on their experience or prior knowledge to seek information in archived software documentation, ignoring unfamiliar and less-used keywords and locations. Moreover, they tend to search only those locations in documents that are either easy to remember or are readily available to them. This strategy, however, results in an inefficient search of the required information, as individuals' (available) experience might not always be consistent with the current state of the software [P13], [P17].

TABLE 9
Effects of cognitive biases

| Bias | Effects | Knowledge area |
|------|---------|----------------|
| Confirmation | Incorrect beliefs about the completeness and correctness of answers or solutions [P13] | Maintenance |
| | Insufficiently exploring code when dealing with a change request [P18] | Construction |
| | Rejecting results and measurements that do not support the analyst's beliefs [P37] | Design |
| | Higher defect rate [P4] | Testing |
| | More post-release defects [P6] | Testing |
| | Misidentifying causes of program failure [P32] | General |
| | Unrealistic trials of SE tools or techniques [P26] | Testing |
| | Running more tests that show a system works than tests that find problems [P31], [P32] | Testing |
| Anchoring and adjustment | Reusing SQL queries introduced errors in the new context [P2] | Construction |
| | Design errors and adding unnecessary functionality [P42] | Design |
| | Reduced long-term productivity [P51] | Management |
| | Ignoring change requests [P17] | Construction |
| | Inaccurate effort estimation [P64] | Design |
| Availability | Ignoring unfamiliar and less-used keywords and locations while searching documentation [P13] | Maintenance |
| | Relying on memorable past experiences that are inconsistent with the current system [P17] | Design |
| | Misrepresenting code features [P32] | General |
| Representativeness | Dropping error-producing test cases [P7] | Quality |
| | Misrepresenting code features [P32] | General |
| Overconfidence | Insufficient effort in requirements analysis [P3] | Requirements |
| | Overall software development failures [P3] | General |
| Hyperbolic discounting | Neglecting refactoring and maintenance for short-term gains [P36] | Design |
| Framing and fixation | Reduced creativity in high-level design [P19] | Requirements |
| Attentional | Misunderstanding UML class diagrams [P28] | Design |
| Miserly information processing | Uncritically agreeing to a client's requirements [P21] | Requirements |
| Bandwagon effect | Uncritically agreeing with a team leader's suggestions [P21] | General |
| Status quo | Illogical defense of previous decisions [P21] | General |

Availability bias is associated with over-representation of memorable code features and certain controversies such as preferring a familiar but less suitable programming language [P32].

### 4.3.5 Fixation and framing effect

The framing effect is the tendency to react differently to situations that are fundamentally identical but presented (i.e., "framed") differently. One primary study argued that framing dubious desiderata as "requirements" leads to undue fixation on the desiderata, consequently reducing design creativity [P19].

### 4.3.6 Attentional bias

Attentional bias refers to how recurring thoughts can bias perception [68]. For example, Siau et al. [P28] investigated experts' interpretations of entity-relationship diagrams with conflicts between their relationships and their labels. For example, one diagram showed a parent having zero-to-many children, when a person obviously needs at least one child to be a parent. The experts failed to notice these contradictions because they only attended to the relationships. The paper claims that this is "attentional bias;" however, it does not explain what "recurring thoughts" were involved, or how they led to ignoring the surface semantics. Therefore, "attentional bias" may not be the best explanation for this phenomenon (see also Section 5.2).

### 4.3.7 Representativeness

Representativeness bias is the tendency to make a judgment by fitting an object into a stereotype or model based on very few properties [36]. Representativeness may explain why "test cases that produce errors may be overlooked ... [which] leads to an increase in software defect density" [P7, p. 5]. Similarly, representativeness may explain why certain "notable features of the code base are judged as frequent, even if they are notable for reasons other than frequency" [P32, p. 59].

### 4.3.8 Miserly information processing, bandwagon effect and status quo bias

Miserly information processing is the tendency to avoid deep or complex information processing [2]. The bandwagon effect is "the propensity for large numbers of individuals, in social and sometimes political situations, to align themselves or their stated opinions with the majority opinion as they perceive it" [39, p. 101]. Status quo bias is the tendency to defend present conditions irrationally [69].

One primary study suggests that: (1) miserly information processing causes professionals to accept client requirements uncritically; (2) the bandwagon effect causes professionals to accept a team leader's decision without considering alternatives; and (3) status quo bias leads the same professionals to defend prior illogical choices [P21]. While plausible, we are not aware of any empirical research that investigates these causal relationships.

### 4.3.9 Overconfidence bias

One primary study suggests overconfidence leads professionals to neglect requirements analysis, leading to a superficial or incorrect understanding of the situation at hand [P3]. Similarly, project managers are often overconfident in their time and resource estimation for projects, which can lead to outright project failure [P3]. However, myriad biases including confirmation bias, miserly information processing and anchoring may contribute to poor understanding of problematic situations and poor estimation (see Section 5.2).

### 4.4 What debiasing approaches are investigated?

One primary study speculates that some techniques—burndown charts, bottom-up planning, product demonstrations, daily team meetings, flexible planning, and stakeholder feedback——may guard against numerous biases [P63]. Others propose specific debiasing techniques (Table 10). While debiasing techniques concentrate on only six of the 37 cognitive biases investigated, some of these may generalize to a larger family of similar or interrelated biases. However, **none** of the primary studies provide strong empirical evidence for the effectiveness of debiasing techniques. All the techniques discussed in this section are untested suggestions.

### 4.4.1 Availability bias

Availability bias often hinders an efficient search for knowledge in software documentation (as discussed in Section 4.3.4). One primary study proposes three techniques to mitigate this problem: 1) developing a "frequently asked questions" document; 2) introducing spelling conventions; and 3) using ontology-based documentation—documents that formalize multiple relationships between discrete pieces of scattered information, to facilitate traversal and search [P13]. Another suggestion for mitigating availability bias is to maintain detailed records of software practitioners' experience gained during software project development lifetime [P53]. (This is unlikely to work because many such documents would be overlong and go out of date quickly.)

Availability bias also affects software design. Project-specific traceability—"*a technique to describe and follow the life of any conceptual or physical software artifact throughout the various phases of software development*" [P17, p. 111], may not only mitigate the effects of availability bias but also anchoring bias and confirmation bias [P17], [P18]. Framing the project context to highlight relevant information may also help to mitigate availability bias during software development and maintenance [P32].

Meanwhile, participatory decision-making and retrospective meetings may ameliorate the effects of availability bias on effort estimation and decision making [P65]. Including project stakeholders and domain experts in retrospective meetings might further mitigate availability bias [P63].

### 4.4.2 Confirmation bias

Despite being one of the most frequently investigated cognitive biases, very few primary studies propose approaches for mitigating confirmation bias. One approach to undermining confirmation bias during testing is to ask developers explicitly to seek evidence of problems rather than evidence that the system functions correctly [P32]. Similarly, asking designers to generate and simultaneously evaluate multiple alternative solutions should mitigate confirmation bias in high-level design [P37]. As discussed in Section 4.4.1, traceability may also reduce confirmation bias [P16].

### 4.4.3 Anchoring and adjustment bias

Anchoring bias is especially problematic during effort estimation, which is a special case of forecasting. Forecasting is an area of intense study in the operations research community; however, this review only covers SE research. Still, forecasting can be divided into expert-based (i.e., based on human judgment) and model-based (i.e., based on a mathematical model). Since anchoring-and-adjustment only occurs in expert-based forecasting, adopting model-based forecasting can be considered a debiasing technique (with the catch that these models are subject to statistical bias and variance) [P53], [P63]. However, model-based forecasting often requires information that is not available in software projects, so several suggestions for debiasing expert-based effort estimation have been proposed.

For example, planning poker—an estimation technique where participants independently estimate tasks and simultaneously reveal their estimates, was proposed to prevent anchoring on the first estimate [P38]. Analogously, generating multiple design candidates may help avoid anchoring on a preconceived solution [P33]. However, the research on group estimation from multiple anchors is mixed (cf. [70] for a summary), and we are not aware of any research investigating how multiple design candidates affect anchoring.

Some primary studies argue that anchoring on an initial estimate can be mitigated by explicitly questioning software project managers via different facilitators (project members or stakeholders). The resulting knowledge sharing should help project managers avoid anchoring their initial estimates on prior experience or self-assessments [P65], [P63].

One study [P37] suggested raising awareness about anchoring bias. While such weak interventions are unlikely to help [45], asking directed questions (discussed next) may be more effective [P3].

### 4.4.4 Overconfidence and optimism bias

Overconfidence and optimism are two of the most investigated cognitive biases in the SE literature (see Fig. 2). Browne and Ramesh [P3] propose addressing several cognitive biases, including overconfidence, using directed questions, which "attempt to elicit information through the use of schemes or checklists designed to cue information in the user's memory." For instance, to reduce anchoring (discussed above) we might ask, "What is your starting point for estimating (that) duration? Why did you start there?"). To reduce overconfidence, we might ask, "Play the devil's advocate for a minute; can you think of any reasons why your solution may be wrong?" [P3].

TABLE 10
Debiasing techniques

| Bias | Debiasing technique | Knowledge area |
|---|---|---|
| Anchoring and adjustment | Traceability [P17], [P18] | Design, Construction |
| | Directed questions [P3] | Requirements |
| | Planning poker [P38] | Management |
| | Generating multiple solutions [P33] | Design |
| | Increasing awareness of the bias, warning participants, disregarding initial anchors or status quo [P37] | Design |
| | Statistical prediction methods [P53] | Process |
| | Technical competence and explicit questioning by facilitators [P65] | Management |
| Availability | Appropriate documentation [P13] | Management |
| | Traceability [P17], [P16] | Design, Construction |
| | Framing information to highlight problematic areas [P32] | General |
| | Maintaining detailed records [P53] | Process |
| | Participatory retrospective decision-making meetings [P65] | Management |
| | Stakeholder feedback and expert involvement [P63] | Management |
| Overoptimism and overconfidence | Directed questions [P3] | Requirements |
| | Planning poker [P62] | Management |
| | Double-loop learning [P45] | Management |
| | Appropriately framing estimation questions [P47] | Management |
| Confirmation | Traceability [P16] | Construction |
| | Explicitly exploring disconfirmatory evidence [P32] | General |
| | Generating multiple solutions [P37] | Design |
| Representativeness | Training, retrospectives, warning about biases, and explicitly asking for alternate solutions [P53] | Process |
| | Maintaining a consistent user interface [P20] | Requirements |

Another primary study argues that Planning Poker helps practitioners mitigate optimism [P62]. However, planning poker does not explicitly encourage converging on more realistic (or pessimistic) estimates. It seems more plausible that the range of estimates would mitigate overconfidence.

Another technique for mitigating overconfidence bias is *double-loop learning* [P45]. This involves encouraging individuals and organizations to engage in self-reflective learning by explicitly confronting initial assumptions and developing more appropriate ones.

Meanwhile, one primary study found that the framing of estimation questions affects optimism. Specifically, asking "How much effort is required to complete X?" provided less optimistic estimates than asking, "How much can be completed in Y work hours?" [P47]. Focusing on tasks (rather than time periods) could therefore be considered a debiasing technique.

### 4.4.5 Representativeness

Several non-evaluated techniques for mitigating the effects of representativeness are proposed, as follows:

- Using consistent user interfaces in decision support systems [P20].
- Explicitly asking practitioners for alternative solutions (or ideas) [P53].
- Encouraging retrospective analysis and assessment of practitioners' judgments [P53].
- Warning practitioners about the bias a priori (which, again, is unlikely to work) [P53].
- Frequently training the participants to avoid representativeness [P53].

TABLE 11
Frequently Used Research Methods

| Research method | # Primary studies |
|---|---|
| Experiment | 30 |
| Case study | 6 |
| Think-aloud protocol analysis | 5 |
| Interview | 3 |
| Grounded theory | 3 |
| Survey | 3 |
| Other approaches | 7 |

### 4.5 What research methods are used?

Table 11 summarizes research methods used most frequently to investigate cognitive biases in SE. Most of the empirical studies employed laboratory experiments. Predominately qualitative research approaches (e.g. case studies) are underrepresented. Nine empirical papers reported multiple studies, of which seven reported multiple experiments while two adopted a multimethodological approach.

*Other approaches* include studies that either used preexisting data sets (e.g. archived data from company records or project documents) or did not report sufficient details about the research method to classify.

### 4.6 When were the articles published?

Fig. 3 shows the number of primary studies published each year. The earliest paper published was in 1990. There were one or two papers per year until 2001, after which we see a noticeable but inconsistent increase, peaking in 2010. The spike seen in 2010 is mostly due to four publications co-authored by Calikli and Bener, who feature in the list of most prolific authors (Section 4.8). Only two of studies were found in 2016. There could be additional 2016 studies that were not indexed at the time of the search (in December, 2016).

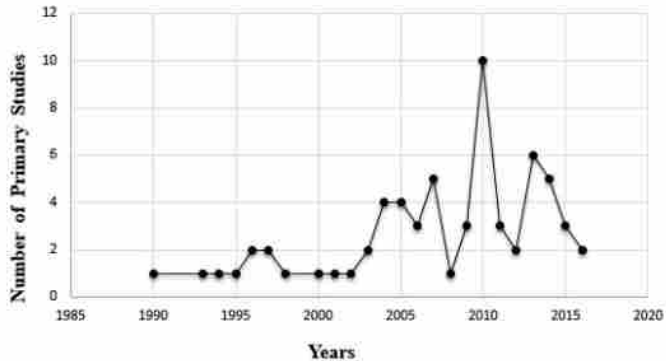

Fig. 3. Publication Trend until 2016

### 4.7 Where were the articles published?

Table 12 shows the most common publication outlets. Articles are scattered across many outlets, with 58% in journals or academic magazines, and the rest in conferences, workshops or symposiums. The outlet with the most articles (5) is Journal of Systems and Software.

### 4.8 Who is conducting the research?

To investigate the most prolific authors, we ranked all 109 authors by the number of publications in the sample. Of these, 81 authors had one paper; 16 authors had two papers; and eight authors had three papers. Four authors stand out with five or more primary studies (Table 13). Magne Jørgensen is clearly the most prolific author working in this area. Some of the authors that feature in the table are frequent collaborators; for instance, Calikli and Bener co-authored eight publications, while Jørgensen and Moløkken collaborated on four.

### 4.9 Which SWEBOK knowledge areas are targeted by research on cognitive biases?

To answer RQ9, we categorized the primary studies using the knowledge areas specified in the Software Engineering Body Of Knowledge [71], as shown in Table 14. The most frequently investigated knowledge area is SE management in 21 primary studies, followed by software construction and software design investigated in 12 and 11 primary studies respectively. Many critical knowledge areas including requirements, design, testing and quality are under-represented.

## 5 DISCUSSION

This section summarizes the implications of this study for SE research, practice and education, followed by its limitations and avenues for future research.

### 5.1 Implications for SE research

The results of this study suggest four ways of improving research on cognitive biases in SE: conducting more qualitative and multimethodological research; investigating neglected areas; better integrating results across studies; and addressing confusion.

First, experiments are clearly the preferred method in our sample (see Section 4.5). This is appropriate insofar as 48 of the primary studies investigate causality, rather than establish facts or develop theories. However, it is crucial to understand not only whether but also how, and how much cognitive biases affect productivity, quality, and software engineering success [72]. Demonstrating a bias in a lab is not the same as establishing a significant effect on real projects. Some biases (e.g. fixation) may completely derail projects (e.g. through complete failure to innovate) while others may have measurable but practically insignificant effects. Measuring effects on specific dependent variables is essential for differentiating high-impact biases from low-impact ones. Additionally, better understanding the mechanisms through which biases sabotage success will help us create effective debiasing practices.

More qualitative and exploratory research is needed to understand where and how cognitive biases manifest in SE practice, and how biases and debiasing techniques are perceived by practitioners. Multimethodological approaches may also help. For example, combining a randomized control trial with a protocol study [73] may help not only demonstrate a causal relationship (experiment) but also reveal the cognitive mechanism underlying the causal relationship (protocol analysis).

Second, many cognitive biases have been investigated in information systems but not software engineering (e.g. ambiguity effect, emotional bias, irrational escalation, reactance, Dunning-Kruger effect [16]). It is plausible that all these biases, and more besides, also affect SE projects. For instance, the Dunning-Kruger effect—the tendency for less skilled people to overestimate their abilities [74]—could explain some conflicts between professionals with different skill levels.

Furthermore, several categories of biases (e.g. social and memory biases) are largely ignored in the SE literature. Similarly, cognitive biases have not been applied to several SE knowledge areas, including configuration management and professionalism (see Table 14). Meanwhile, 21 of the cognitive biases reviewed here were investigated by only one or two studies. Most of the primary studies simply demonstrate biases; few propose debiasing techniques (e.g. [P18], [P45]), which are crucial for practical impact.

Third, research on cognitive biases in SE suffers from the same problems as research on cognitive biases in psychology: most studies investigating or discussing the same phenomenon are disconnected from each other (see Section 5.5). Cognitive phenomena rarely act in isolation. Better integrating across studies, biases and knowledge areas may be necessary to understand complex cognitive explanations for routine SE problems. For example, design creativity is inhibited by the nexus of framing effects, fixation and requirements engineering practices [3].

Fourth, research on cognitive biases in SE (and to a lesser extent in cognitive psychology) suffers from widespread confusion, which we address in the next subsection.

TABLE 12
Publication Outlets

| Source Avenue | # Primary Studies |
|---|---|
| Journal of Systems and Software | 5 |
| Journal of Software | 3 |
| Systems Engineering Conference | 3 |
| International Conference on Software Engineering | 3 |
| Information and Software Technology Journal | 2 |
| Empirical Software Engineering Journal | 2 |
| International Conference on Evaluation and Assessment in Software Engineering | 2 |
| IEEE Transactions on Software Engineering | 2 |
| Communications of the ACM | 2 |
| Information Systems Journal | 2 |
| International Conference on Information Systems | 2 |
| International Conference on Predictive Models in Software Engineering | 2 |
| International Symposium on Empirical Software Engineering and Measurement | 2 |
| Psychology of Programming Interest Group Workshop | 2 |
| Information & Management Journal | 1 |
| Advanced information systems engineering Journal | 1 |
| International Journal of Project Management | 1 |
| European Conference on Cognitive Science | 1 |
| AI Magazine | 1 |
| Americas Conference on Information Systems | 1 |
| Canadian Conference on Electrical and Computer Engineering | 1 |
| Document Engineering (Symposium) | 1 |
| European Conference on Information Systems | 1 |
| European Software Process Improvement | 1 |
| IASTED International Conference on Software Engineering | 1 |
| Information Systems Management Journal | 1 |
| Information Systems Research Journal | 1 |
| International Conference on Electronics, Communications and Computers | 1 |
| International Conference on Information Science and Technology | 1 |
| International Conference on Information, Business and Education Technology | 1 |
| International Conference on Intelligent user interfaces | 1 |
| International Workshop on Emerging Trends in Software Metrics | 1 |
| Journal of Association for Information Systems | 1 |
| Journal of Visual Languages | 1 |
| Management Science Journal | 1 |
| Proceedings of AGILE conference | 1 |
| Research in Systems Analysis and Design: Models and Methods Journal | 1 |
| Scandinavian Journal of Information Systems | 1 |
| Science of Computer Programming Journal | 1 |
| Software Quality Journal | 1 |
| Transaction on Professional Communication Journal | 1 |
| Ubiquity Journal | 1 |
| Workshop on a General Theory of Software Engineering (GTSE) | 1 |
| International Workshop on Cooperative and Human Aspects of Software Engineering | 1 |
| International Conference on Human-Computer Interaction | 1 |

TABLE 13
Cognitive biases investigated by top SE researchers

| Researcher | # Primary studies | Cognitive biases investigated | Knowledge area |
|---|---|---|---|
| Jørgensen | 13 | Optimism, confidence, anchoring and adjustment, hindsight, wishful thinking, impact, selective perception, representativeness, availability, misleading information | Management, construction, general |
| Bener | 8 | Confirmation, representativeness, availability, anchoring and adjustment | Construction, testing |
| Calikli | 8 | Confirmation, representativeness, availability, anchoring and adjustment | Construction, testing |
| Moløkken | 5 | Confidence, anchoring and adjustment, optimism | Management, construction, general |

TABLE 14
SE Knowledge Areas Investigated

| Knowledge area | # Primary studies |
| --- | --- |
| SE Management | 21 |
| Software Construction | 12 |
| Software Design | 11 |
| Software Requirements | 7 |
| Software Testing | 7 |
| SE (General) | 3 |
| Software Quality | 1 |
| Software Maintenance | 1 |
| SE Process | 1 |
| SE Models & Methods | 1 |
| Engineering Foundations | 0 |
| Software Configuration Management | 0 |
| SE Economics | 0 |
| SE Mathematical Foundations | 0 |
| SE Professionalism | 0 |

## 5.2 Widespread confusion

Analyzing many of the primary studies was hindered by several interconnected problems.

Some biases are defined inconsistently between studies. Some biases have multiple names (e.g. system justification is the same as status-quo bias). Some ostensibly different biases are extremely similar, or likely just the same underlying cognitive process manifesting in different circumstances. For example, when we get preoccupied with a number, it is called *anchoring bias* [22]. But when we get preoccupied with a default option, it is called *default bias* [69]. We call preoccupation with presentation the *framing effect* [2]. Preoccupation with information that supports our beliefs is *confirmation bias* [75], while preoccupation with the typical use of an object is *functional fixedness* [76]. Preoccupation with the first few items in a list is *primacy*, while preoccupation with current conditions is *status quo bias*. If we better understood attention and perception, we might not need all these terms to explain what people attend to.

Meanwhile, the primary studies reviewed here attempt to apply concepts from a social science discipline to an applied science context, which is often quite different from the context in which those concepts originated. Communication across scientific communities can be difficult, and some miscommunication is normal [77].

These problems lead to several points of confusion:

1) Dubious causal relationships between biases. For instance, arguing that framing causes fixation [3] is questionable because both the 'cause' and 'effect' may be the same phenomenon in different guises. One primary study addressed this by organizing similar biases into "biasplexes" [P22] to help reason about debiasing.

2) Confusing biases with heuristics and fallacies. Cognitive biases are patterns of errors. *Heuristics* are simple but imperfect rules for judging and deciding [36]. Heuristics like anchoring-and-adjustment *cause* biases like adjustment bias when the heuristic leads to a pattern of errors under certain circumstances.

Meanwhile, fallacies are *specific* errors in logic [78]. For example, the sunk cost fallacy is a logical error in which past (lost) investment is used to justify further (irrational) investment [79]. Some cognitive biases (e.g. sunk cost bias) are basically the tendency for different people to make the same logical fallacy under similar circumstances.

3) Some articles simply use terms incorrectly. For instance, one article claimed that the "Parkinson's Law Effect' is a cognitive bias; specifically, the tendency to procrastinate until the day a task is due. As far as we can tell, there is no such effect. Parkinson's Law, which is not a cognitive bias, claims that "work expands so as to fill the time available for its completion." [80]. It means that people keep working on things until they are due, not that they wait until the due date to start. Other articles seemed to misunderstand attentional bias, confirmation bias and many others.

To address these areas of confusion, we offer suggestions to both researchers and editors. For researchers, applying theories from reference disciplines to new contexts is often difficult. Skimming a single article or Wikipedia entry is not sufficient to understand a cognitive bias or psychological theory. Depending on the theory and researcher, it may be necessary to take courses in cognitive psychology or conduct an extensive literature review. Finding a mentor or collaborator with a strong background in the bias or theory is also wise. (We discussed aspects of this paper with three different professors of psychology to make sure that our interpretations are reasonable and our nomenclature correct). Meanwhile, editors faced with a manuscript applying a theory from a reference discipline to SE should consider inviting a review from someone with a strong background in that theory or reference discipline, even if they are in another discipline. This practice can be modeled after statistical review in medical journals (cf. [81]).

## 5.3 Implications for SE practice

The most important takeaways of this research for practitioners are as follows:

1) Cognitive biases are universal human phenomena. They affect everyone and likely disrupt all aspects of software development, regardless of industry, platform, programming language or project management philosophy.

2) Tasks are often easier to debias than people [45]. Although, awareness can help people avoid situations in which the biases arise [82], many debiasing practices can mitigate biases quickly and economically. Learning to perceive and resist anchoring and adjustment is difficult. Planning poker is easy. Learning to write judicious unit tests takes years, but having a whole team spend a day trying to break a system to win a gift basket works instantly. Trying to teach a product designer to be more creative is daunting, but a team can run through a series of lateral thinking exercises in an afternoon.

3) None of the research reviewed here provides compelling empirical evidence that any specific debiasing approach will be effective in a specific future project. Debiasing techniques may have unintended consequences. Caution is warranted.

Some debiasing techniques that have been proposed for software engineering contexts, but have not been extensively validated, include:

1) Planning poker to prevent anchoring bias in effort estimation [P38].
2) Reference class forecasting or model-based forecasting to mitigate optimism in effort estimation [P45].
3) Ontology-based documentation to mitigate availability bias [P13].
4) Directly asking for disconfirmatory evidence to reduce confirmation bias [P32].
5) Generating multiple, diverse problem formulations and solution candidates to mitigate framing effects [P53].
6) Using confirmation bias metrics to select less-biased testers [P6].
7) For security, employing independent penetration testing specialists to reduce confirmation bias [83].
8) Designating a devil's advocate in retrospective meetings to avoid groupthink [84].

### 5.4 Implications for SE education

Reasoning can be debiased through extensive training [45]. What would constitute extensive training varies depending on the bias.

For example, extensive training in effort estimation might consist of estimating thousands of user stories, with immediate and unambiguous feedback. This is probably impractical, especially considering the great variety of software projects and the domain knowledge necessary to estimate tasks accurately.

In contrast, extensive training in design creativity might be more practical. For example, consider an assignment where students have to design a suitable architecture for a given teaching case using UML. Now suppose students are instead asked to design four alternative architectures, all of which should be reasonable and quite different from each other. Or, suppose students are asked to design just two architectures, one uncontroversial and one radical. If most design-related assignments throughout their degrees call for multiple alternatives, generating several design candidates will seem natural. Students are thus more likely to carry this practice into their professional work, much like the programming styles and testing practices taught today. Implementing all the alternatives is unnecessary.

Whether and how to teach the theory of cognitive biases or particular debiasing techniques is another matter. Rather than teaching it as a stand-alone topic, it may be more efficient to explain many cognitive biases in the context of common problems and practices.

For example, a common topic in software project management courses is effort estimation. The instructor shares empirical research on poor effort estimation and the damage it can cause, and explains how Scrum and Extreme Programming prescribe planning poker. The students might then do a planning poker exercise on their term project. This presents planning poker through a method lens. In contrast, the instructor can present it through a theory lens. First, we explain the problem (which is readily understood), followed by the key causes of the problem: 1) anchoring bias; 2) optimism bias; 3) pressure from management and stakeholders to deliver faster. Then, we describe specific debiasing techniques: planning poker for anchoring bias and reference class forecasting for optimism bias. This approach gives students a more nuanced, evidence-based conception of both the problem and potential solutions.

### 5.5 Limitations

The findings of this research should be considered in light of several limitations.

The concept of cognitive biases is not without problems: "the explanations of phenomena via one-word labels such as availability, anchoring, and representativeness are vague, insufficient, and say nothing about the processes underlying judgment" [85, p. 609]. Most cognitive biases "have been studied in complete isolation without any reference to the other ones ... thus resembl[ing] an archipelago spread out in a vast ocean without any sailor having visited more than one (or at the most two) of the small islands yet." [86, p. 399]. Cognitive biases do not constitute a comprehensive theory of reasoning; they are more like a set of empirical generalizations regarding common reasoning problems. Except, many of the problems seem related, and both their relationships and causes are mysterious. Synthesizing research based on cognitive biases is therefore epistemologically fraught.

Furthermore, inconsistent terminology in the primary studies, as discussed in Section 5.2, impedes analysis. The analyst must constantly question whether what is said is what is meant. This extra layer of interpretation creates additional possibilities for errors.

Our search string may have missed relevant studies published in computer science outlets. Further relevant studies may not be indexed by the databases we searched. Limiting our search to 'computer science' or similar categories (see Table 2) may have excluded relevant studies in other domains (e.g. a study of a cognitive bias in an SE context published in a psychology journal). ' We used several tactics to improve external validity, as explained in Section 3.5):

1) we employed backward snowballing on reference lists of primary studies;
2) we analyzed the profiles of most prolific authors to check if we have missed any relevant studies;
3) being well versed with this topic, we discussed as a group whether we knew of any relevant studies that were suspiciously absent from sample;
4) we analyzed the first 150 results of the same search on Google Scholar;
5) we searched the for the names of some specific cognitive biases.

Despite these precautions, however, some relevant studies may still have been missed. In particular, because we excluded unpublished works; the results are subject to publication bias.

### 5.6 Future research

The research reviewed here does not support many concrete, practical recommendations for overcoming cognitive biases in real software projects. Better recommendations will require several parallel lines of inquiry:

1) understanding the social and cognitive processes (in software development) from which cognitive biases emerge;
2) measuring the effects of specific biases in specific contexts to see what's practically important;
3) adapting existing debiasing approaches for software engineering contexts, and developing new ones;
4) empirically evaluating the effectiveness of debiasing techniques.

Each of these lines of inquiry require different kinds of studies. Qualitative studies, including case studies [87], grounded theory [88] and protocol studies [89], [90] are well suited to understanding social and cognitive processes. Meanwhile, because software engineering success is so multifaceted [91], sophisticated statistical modelling is needed to estimate the effects of a specific bias. Creating new debiasing approaches and adapting existing ones for software contexts is engineering research in the sense that the research seeks to invent a useful artifact. The debiasing artifacts can be evaluated using controlled experiments [92] or action research [93].

Meanwhile, we need a better classification of cognitive biases. We found that some biases are difficult to classify in Fleischmann et al.'s [16] taxonomy—some appear related to multiple categories; others do not fit in any category (Section 4.1). Better organizing cognitive biases according to their causal mechanisms, the nature of their effects or their specific manifestations in software projects would facilitate reasoning about biases and possibly inform debiasing techniques.

## 6 CONCLUSION

This study provides a comprehensive view of research on cognitive biases in software engineering. It identifies 65 primary studies exploring 37 cognitive biases, organized into eight categories. The most investigated cognitive biases belong to *interest* and *stability* biases, whereas, the class of *social* and *decision* biases are least investigated. A constant and sustained interest in this area is clear, especially in software project management. Substantial opportunities for impactful research remain not only in other areas (e.g. design, testing, and devops) but also considering other biases (e.g. selective perception, input bias and priming effect).

Some of the key findings of this paper are:

- SE research focuses on multiple antecedents and effects of cognitive biases; however, the psychological and sociological mechanisms underlying many cognitive biases are poorly understood.
- We found debiasing approaches for only six cognitive biases. None were empirically evaluated and few were grounded in psychological theory.
- Controlled experiments are the most common research method for investigating cognitive biases in

SE. However, we found no replications, families of experiments or meta-analyses.
- Most of the primary studies are disconnected. That is, many studies do not refer to previous SE research on the same bias, or conceptualize the bias in the same way.

In summary, cognitive biases help to explain many common problems in SE. While they are robust against simply raising awareness, professionals can often be debiased through simple interventions. The better we understand the theoretical foundations of common problems and practices, the easier it will be to develop effective interventions to mitigate biases and therefore alleviate their corresponding problems. Ultimately, we hope that this literature review will encourage many SE researchers to further explore the phenomena of cognitive biases in SE and thus will serve as a basis for future research.

## ACKNOWLEDGMENTS

The authors would like to thank the associate editor and three anonymous reviewers for their suggestions on earlier versions of this paper. This study was supported in part by the Infotech Oulu Doctoral grant at University of Oulu to Rahul Mohanani and Iflaah Salman.

## APPENDIX A
## DEFINITIONS OF COGNITIVE BIASES INVESTIGATED IN SE

| Cognitive bias | Category | Definition | Primary studies |
|---|---|---|---|
| Anchoring and adjustment bias | Stability | "The tendency, in forming perceptions or making quantitative judgments of some entity under conditions of uncertainty, to give excessive weight to the initial starting value (or anchor), based on the first received information or one's initial judgment, and not to modify this anchor sufficiently in light of later information" [39, p. 51]. Adjustment bias is a psychological tendency that influences the way people intuitively assess probabilities [36]. | P1, P2, P7, P10, P11, P16, P17, P18, P20, P22, P30, P33, P34, P37, P38, P42, P43, P47, P49, P51, P53, P56, P57, P60, P63, P64, P65 |
| Attentional bias | Perception | The tendency of our perception to be affected by our recurring thoughts [68]. | P28, P40 |
| Availability bias | Pattern-recognition | Availability bias refers to a tendency of being influenced by the information that is easy to recall and by the information that is recent or widely publicized [40]. | P7, P10, P11, P13, P16, P17, P18, P23, P29, P32, P37, P39, P43, P60, P63, P64, P65 |
| Bandwagon effect | Social | "The tendency for large numbers of individuals, in social and sometimes political situations, to align themselves or their stated opinions with the majority opinion as they perceive it" [39, p. 101]. | P21, P22 |
| Belief perseverance | Stability | "The tendency to maintain a belief even after the information that originally gave rise to it has been refuted or otherwise shown to be inaccurate" [39, p. 112]. | P22 |
| (Over) confidence bias | Action-oriented | The tendency to overestimate one's own skill, accuracy and control over one's self and environment [62]. | P3, P16, P20, P37, P41, P45 |
| Confirmation bias | Interest | The tendency to search for, interpret, focus on and remember information in a way that confirms one's preconceptions [94]. | P4, P5, P6, P7, P8, P9, P10, P12, P13, P14, P16, P17, P18, P20, P21, P26, P27, P30, P31, P32, P36, P37, P39, P43, P44, P55, P64 |
| Contrast effect | Perception | The enhancement or reduction of a certain perception's stimuli when compared with a recently observed, contrasting object [95]. | P36 |
| Default bias | Stability | The tendency to choose preselected options over superior, unselected options [69]. | P22 |
| Endowment effect | Stability | "The tendency to demand much more to give up an object than one is willing to pay to acquire it" [96]; i.e., when someone values something more just because they own it. | P25 |
| Fixation | Pattern-recognition | The tendency to disproportionately focus on one aspect of an event, object, or situation, especially self-imposed or imaginary obstacles [97]. | P19 |
| Framing effect | Perception | "The tendency to give different responses to problems that have surface dissimilarities but that are really formally identical" [2] | P19, P20, P64 |
| Halo effect | Perception | The halo effect can be defined as the tendency to use global evaluations to make judgments about specific traits [65]. | P35, P63 |
| Hindsight bias | Memory | When people know the actual outcome of a process, they tend to regard that outcome as having been fairly predictable all along—or at least more predictable than they would have judged before knowing the outcome [63]. | P49, P57, P60, P63 |
| Hyperbolic discounting | Decision | The tendency of people to prefer options that offer smaller rewards with more immediate pay-off to options with larger rewards promised for future [67]. | P36 |
| IKEA effect / I-designed-it-myself effect | Interest | The tendency for people to ascribe greater value to items that they have created, designed or assembled [62]. | P25 |
| Impact bias | Action-oriented | Impact bias is the tendency for people to overestimate the length or the intensity of future feeling states [98]. | P49, P57 |
| Information bias | Other | The tendency to request unnecessary or unhelpful information, especially in times of uncertainty [66]. | P10, P15, P36 |
| Infrastructure bias | Decision | The location and availability of preexisting infrastructure such as roads and telecommunication facilities influences future economic and social development [99]. | P37 |
| Invincibility bias | Action-oriented | The tendency to over trust one's own abilities [99]. | P37 |
| Mere exposure effect | Pattern recognition | "Increased liking for a stimulus that follows repeated, unreinforced exposure to that stimulus" [100, p. 31]. | P22, P63, P65 |
| Miserly information processing | Action-oriented | The tendency to avoid deep or complex information processing [2]. | P21 |
| Misleading information | Action-oriented | The tendency to blindly follow provided information without being able to self-evaluate [101]. | P21, P49, P57 |
| Neglect of probability | Decision | The tendency to disregard probability during decision-making [102]. | P36 |
| Normalcy effects | Action-oriented | The tendency to systematically "underestimat[e] the probability or extent of expected disruption" during a disaster" [59]. | P22 |
| (Over) optimism bias | Action-oriented | The tendency to be over-optimistic, overestimating favorable and pleasing outcomes [102]. | P22, P24, P31, P45, P46, P48, P50, P51, P52, P54, P57, P58, P59, P61, P62 |
| Primacy and recency effects | Perception | The tendency to remember the first and last few items in a sequence more than those in the middle [67]. | P36 |
| Representativeness | Other | The tendency to reduce many inferential tasks to simple similarity judgments [36]. | P7, P29, P32, P43, P53, P60 |
| Selective perception | Perception | The tendency for different people to perceive the same events differently. [95]. | P60 |
| Semantic fallacy | Perception | The tendency to pay less attention to the semantics of model than to its syntax or structural constrains (especially when they conflict) [103]. | P40 |
| Semmelweis reflex | Pattern recognition | Unthinking rejection of new information that contradicts established beliefs or paradigms [104]. | P22 |
| Status-quo bias / System justification | Stability | The tendency to irrationally prefer, maintain and defend current conditions, operating procedures, status, or social order [105]. | P21, P22, P37 |
| Sunk cost bias | Decision | Sunk cost bias is the tendency to invest more future resources in a situation in which a prior investment has been made, compared to a similar situation in which a prior investment has not been made [66]. | P63 |
| Time-based bias | Memory | Time-based bias involves a reduction of attention via short-term thinking and hyperbolic discounting errors [106], [107]. | P30, P36 |
| Valence effect | Interest | The tendency to give undue weight to the degree to which an outcome is considered as positive or negative when estimating the probability of its occurrence [108]. | P22 |
| Validity effect | Interest | "The validity effect occurs when the mere repetition of information affects the perceived truthfulness of that information, and appears to be based on recognition memory and may be an automatic process. The validity effect occurs similarly for statements of fact that are true and false in origin, as well as political or opinion statements" [109, p. 211]. | P22 |
| Wishful thinking | Interest | The tendency to underestimate the likelihood of a negative outcome and vice versa [40]. | P22, P49, P57 |

# APPENDIX
# B: LIST OF PRIMARY STUDIES